\documentclass[twocolumn,pra,showpacs]{revtex4-1}
\usepackage{amsmath}
\usepackage{graphics}
\usepackage{epsfig}

\usepackage{color}
\usepackage{ulem}

\newcommand{\Wcm}{\,\mathrm{W/cm}^2}

\newcommand{\Eip}{E_\mathrm{IP}}

\usepackage{ulem}

\begin{document}

\title{Momentum-resolved study of the saturation intensity in multiple ionization}

\author{P. Wustelt$^{1,2}$}
\author{M. M\"oller$^{1,2}$}
\author{T. Rathje$^{1}$}
\author{A. M. Sayler$^{1,2}$}
\author{T. St\"ohlker$^{2}$}
\author{G. G. Paulus$^{1,2}$}

\affiliation{${}^1$Institute of Optics and Quantum Electronics, Friedrich Schiller University Jena, Max-Wien-Platz 1, 07743 Jena, Germany}
\affiliation{${}^2$Helmholtz Institut Jena, Fr\"obelstieg 3, 07743 Jena, Germany}

\begin{abstract}
We present a momentum-resolved study of strong field multiple ionization of ionic targets. Using a deconvolution method we are able to reconstruct the electron momenta from the ion momentum distributions after multiple ionization up to four sequential ionization steps. This technique allows an accurate determination of the saturation intensity as well as of the electron release times during the laser pulse. The measured results are discussed in comparison to typically used models of over-the-barrier ionization and tunnel ionization.
\end{abstract}

\maketitle

Atoms exposed to super-intense laser pulses can be ionized to high charge states. In the optical regime, the ionization probability depends highly nonlinear on the field strength. Therefore, for a pulsed field, ionization is concentrated in a narrow intensity and a correspondingly narrow time interval for each ionization step: There is virtually no ionization at lower intensity because of negligible ionization rate and no ionization at higher intensity because all atoms or ions of the respective charge state are already ionized.  Accordingly, the intensity where ionization peaks is an important quantity for many research areas of intense laser matter interaction. In the literature, different definitions and names are used for the characterization for this or related phenomena (e.g. appearance intensity)~\cite{Augst1989,Chang1993}. We will use the term saturation intensity for this quantity. This is of particular interest, e.g. in laser plasma physics, where correct modeling of ionization over a large number of charge states, which determines the spatio-temporal plasma density, is essential~\cite{Chen2013a}. 

The accurate determination of the saturation intensity is an important problem for modeling strong-field laser matter interaction. Difficulties exist on both on the experimental and the theoretical side. For the latter, a frequently used approach for the estimation of the saturation intensity is the over-the-barrier
ionization (OBI)~\cite{Augst1989}, i.e. the intensity where the Coulomb barrier is lowered below the ground state. Another approach is based on the tunneling rate, i.e. the theories of Perelomov, Popov and Terentev (PPT)~\cite{Perelomov1966} or Ammosov, Delone and Krainov (ADK)~\cite{Ammosov}. Both approaches have inherent problems: For OBI, the entire concept is controversial~\cite{Shakeshaft1990,Pfeiffer2011b}, for tunneling it is known that the approximations required to derive the respective formulas are not valid at the intensities typical for saturation with femtosecond pulses~\cite{Ilkov1992,Scrinzi1999,Tong2005}. Experimentally the problem has been investigated by measuring the intensity dependence of the charge resolved ion yield~\cite{Yamakawa2004,Palaniyappan2005}, thereby the reliable determination of the intensity is the major problem. A meaningful verification of the predictive power of one or the other theoretical model is non-trivial, since the actual field strength at which a specific ionization event occurred is, in general not accessible experimentally.

In this article, we communicate an experiment and method of data evaluation that enables an in situ intensity measurement of the saturation intensity for ions of various charge states simultaneously. A dense beam of Ne$^+$ ions is ionized up to charge state 5 by an elliptically polarized laser field and the full 3D momentum of the ionized particles is measured. The interpretation is analogous to the so-called attoclock technique~\cite{Maharjan2005,Eckle2008a,Fleischer2011}: the vast majority of ionization events takes place at times $t_0$ when the laser field is at the major axis positions of the polarization ellipse. A photoelectron released at $t_0$ will subsequently acquire a drift momentum $p=eA(t_0)$, i.e. the momentum distribution of the photoelectrons is given by the laser field's vector potential $A$ at the instant of ionization and thus can be used to determine the intensity at this time. Errors due to the approximations used to derive $p=eA(t_0)$ are small for $|p|$ \cite{Boge2013}($\approx$1.5$\%$ for the first ionization step and even less for the higher charge states), in particular for the conditions of small Keldysh parameters encountered here. 
The challenge is rather to determine the photoelectrons' momenta when several electrons are emitted within a few femtoseconds.
\begin{figure*}[htbp]
\includegraphics[width=1\textwidth]{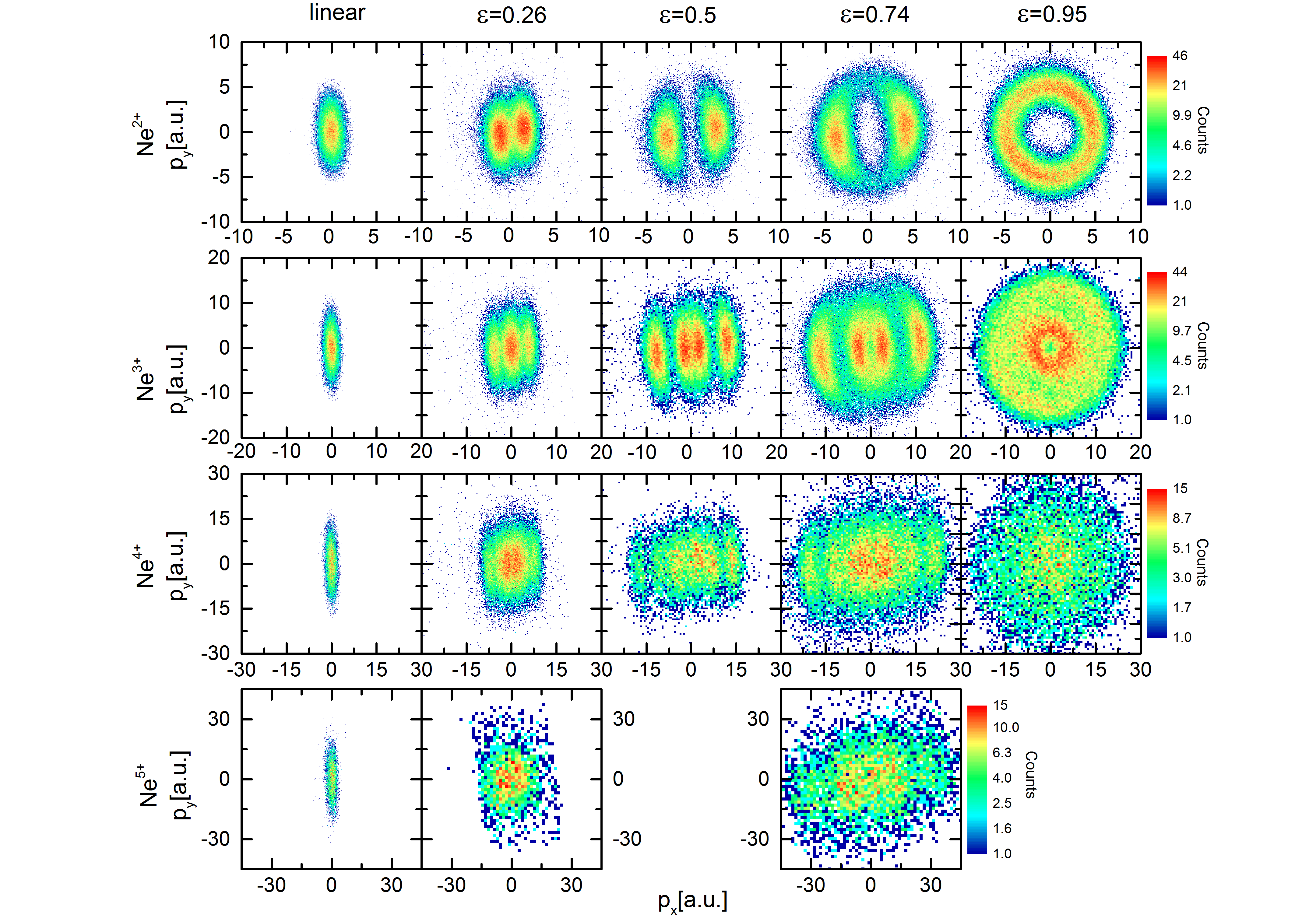}
  \caption{(color online). Measured ion momentum distributions in the polarization plane for different ellipticities (columns) from linear to nearly circular polarization for single, double, triple and quadruple ionization (rows) of Ne$^+$ ions. The momentum distributions of each column are recorded simultaneously. The major polarization axis is parallel to the $y$ axis. Note, a.u. denotes atomic units. The scale is different for the different rows. In each panel it is the same for $p_x$ and $p_y$. The peak intensity of the laser pulse is $\approx 10^{17}\,\Wcm$. Due to small count rates in the quadruple ionization only three ellipticities were measured with significant statistics.}
	\label{fig:dis_neon}
\end{figure*}

In the experiment, the beam of Ne$^+$ ions is produced in a hollow-cathode discharge duoplasmatron ion source~\cite{Lejeune1974} and accelerated to an energy of 8\,keV. Intensities of up to about $10^{17}\Wcm$ are achieved in the interaction region using 10-mJ, 35-fs pulses at a repetition rate of 1\,kHz from a commercial tabletop Ti:Sapphire laser system. The ellipticity, $\varepsilon$, is adjusted by a quarter-wave plate. The three-dimensional momentum distributions are reconstructed from the time and position information recorded for each ion by a delay-line detector \cite{Rathje2013}. 

The ion momentum distributions measured for different ellipticities in the polarization plane of the laser for the cases of single, double, triple and quadruple ionization of Ne$^+$  are shown in Fig.~\ref{fig:dis_neon}. For single ionization with linear polarization one observes a two dimensional Gaussian distribution with its maximum at zero momentum. As expected, the width of the distribution is larger in polarization direction than perpendicular to it. For increasing ellipticities, the single maximum of the distribution splits into two along the minor axis of the polarization ellipse ($x$ axis). At even higher ellipticity, the two peaks become less distinct until a ring with almost constant count rate in tangential direction is obtained for nearly circular polarization. Due to momentum conservation, the momentum distributions of ions and electrons are just mirror images of each other for single ionization.
  
It is well known that these observations can be understood qualitatively and even quantitatively by the so-called simple-man's model (SMM)~\cite{H.B.vanLindenvandenHeuvell1987}. Conservation of canonical momentum immediately yields the already cited relation $p=eA(t_0)$. The shape of the momentum distribution follows the geometry of the vector potential while the ionization probability, i.e. the count rate, follows the instantaneous electric field strength \cite{Corkum1989,Litvinyuk2003}. Therefore, the radial size of the distributions is related to the vector potential at the instant of ionization. Conversely, if the pulse shape is known, the radius can be used to determine the ionization time. This is referred to as the hour-hand of the attoclock. The angular offset of the centroid of the distribution with respect to the minor polarization axis is known as the minute-hand \cite{Maharjan2005}. It plays no role in this work. 

\begin{figure}[htbp]
\includegraphics[width=1\columnwidth]{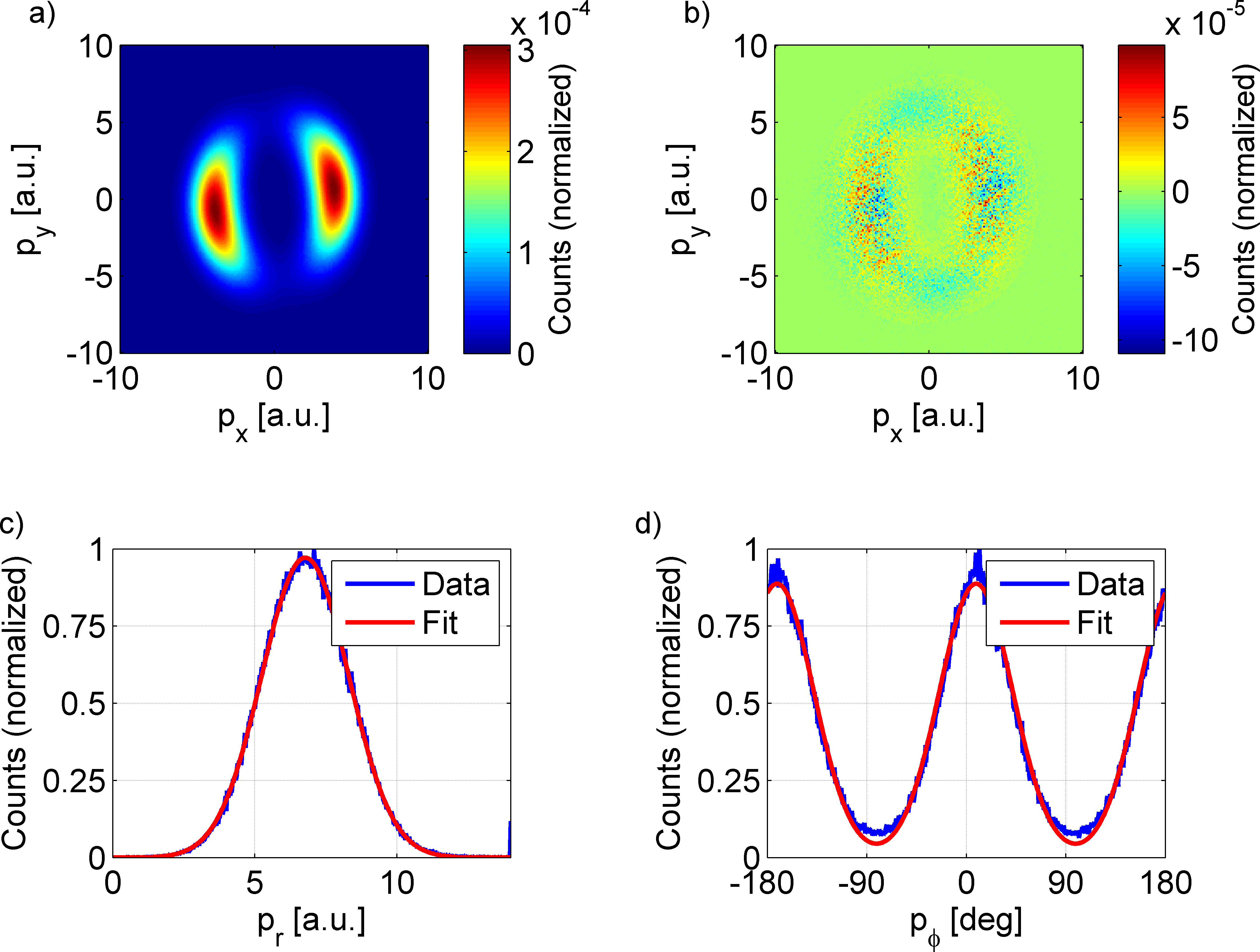}
  \caption{(color online) Single ionization fit function: a) fit result for measured ion momentum with $\epsilon=$0.74 (see Fig. \ref{fig:dis_neon}) , b)  difference of measurement and  fit result, c) projection of data on radial coordinate $p_r$ with fit, d) projection of data on on angular coordinate $p_{\phi}$ with fit.}
	\label{fig:fig_supp1}
\end{figure}

In order to quantify the observations and to subsequently enable the determination of photoelectrons' momenta for each ionization step, we fit a two-dimensional Gaussian distribution, $F$, in polar coordinates $p_r$ and $p_{\phi}$ corrected for ellipticity \cite{Pfeiffer2011a} to the data:
\begin{equation}
F_{\rm single}(p_r,p_{\phi}) := \frac{1}{p_r} e^{ -\left(\frac{p_r-p_{r0}}{\Delta p_r }\right)^2} \cdot e^{ -\left(\frac{p_{\phi}-p_{\phi 0}}{\Delta p_{\phi}}\right)^2 } .
\label{eq:gaus}
\end{equation}
Note that $F_{\rm single}$ represents a two-dimensional fit function, where $p_{r0}$ is the mean radial momentum, and $p_{\phi 0}$ is the angular rotation offset  of the single ionization distributions measured along the vector potential of the ionizing laser field. The factor $1/p_r$ accounts for the volume element in polar coordinates. The widths of the momentum distributions along the ellipticity-corrected coordinates are described by $\Delta p_r$ and $\Delta p_{\phi}$. The function fits very well and enables a concise parametrization of the the measured data with four quantities~(see Fig.\ref{fig:fig_supp1}).

The second row of Fig.~\ref{fig:dis_neon} shows the Ne$^{3+}$ momentum distributions, i.e. the result of double ionization of Ne$^{+}$. For sequential multiple ionization, the resulting momentum of the ion core is the sum of the momenta gained from each of the ionization steps, i.e. it is the convolution of several single ionization steps~\cite{Maharjan2005,Wang2012}. For double ionization, the two photoelectrons can be emitted in the same or in opposite directions such that four peaks emerge. If both electrons are emitted at the similar intensity, the magnitude of their momentum will be similar. Then, the two inner peaks will overlap at zero momentum such that only three peaks are visible, see e.g. Fig.~\ref{fig:dis_neon} b).

\begin{figure}[htbp]
\includegraphics[width=1\columnwidth]{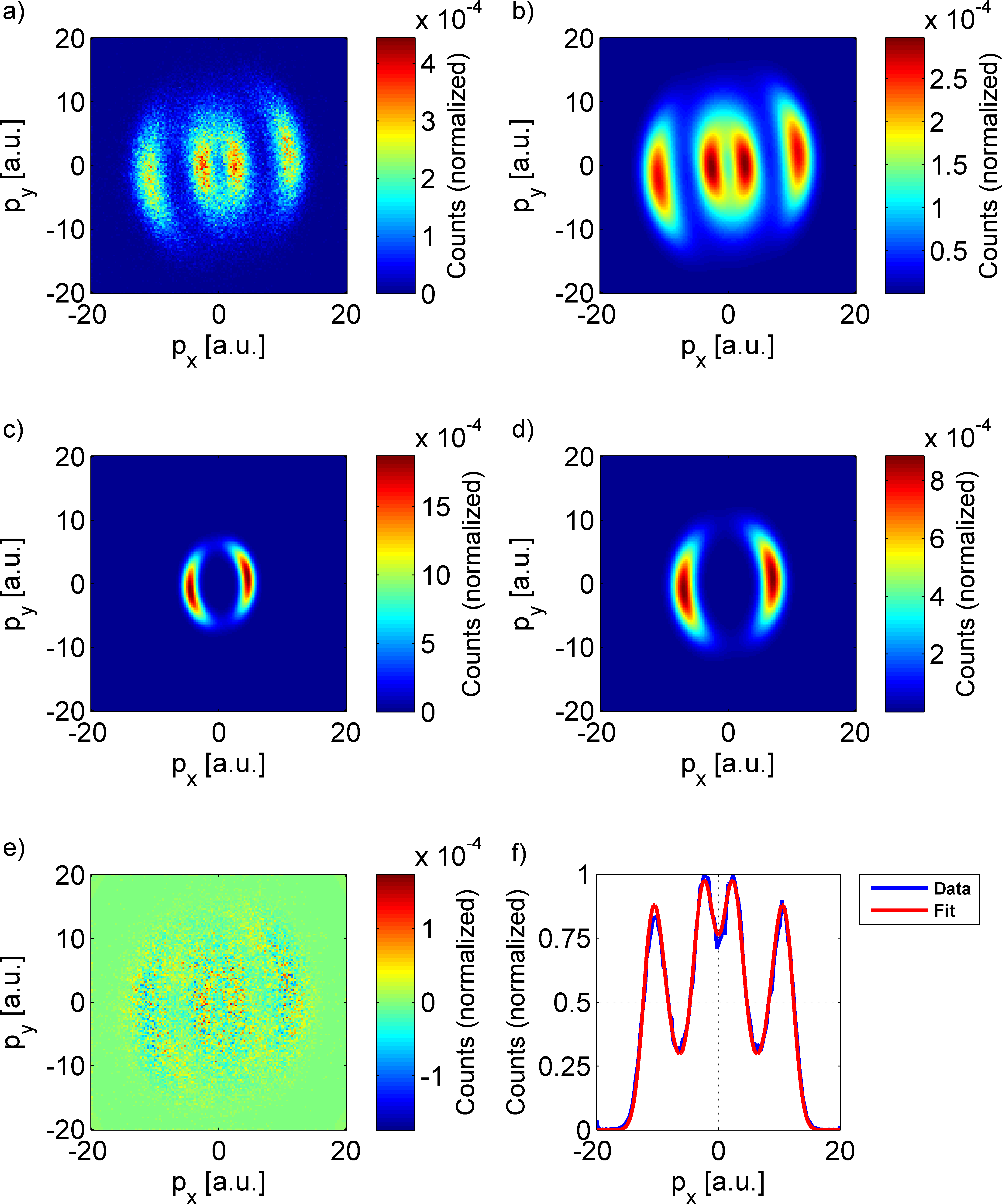}
 \caption{(color online) a) Measured ion distribution of Ne$^{3+}$ for ($\epsilon=$0.74) b) Simulated momentum distribution using the parameters obtained from fitting $F_{\rm double}$ to the measured distribution  c) Simulated momentum distribution for $F_{\rm single 1}$, i.e. the distribution of the first emitted electron, using the fit result d) Simulated momentum distribution for $F_{\rm single 2}$, i.e. the distribution of the second emitted electron, using the fit result, e) Difference of simulated and measured momentum distribution (See Fig. 1). Statistically, the result cannot be distinguished from noise. f) projection of data on $p_x$ with fit.}
\label{fig:Deconvolution}
\end{figure}
The third and forth row of Fig.~\ref{fig:dis_neon} show the results of triple and quadruple ionization, i.e. the distributions of Ne$^{4+}$ and Ne$^{5+}$. Similar to double ionization, the momentum distribution exhibits a multiple peak structure. According to the explanation above, an eight- and 16-peak structure is expected due to the different combinations of emission directions for the emitted electrons. Again, the number of observable peaks will depend on the difference of the photoelectron momenta corresponding to subsequent ionization steps. In general, this difference will be the larger, the larger the difference of the ionization potentials for subsequent charge states are, i.e. the larger the differences of the intensities at which the ionization events take place. For Ne$^{5+}$, the visibility of the peaks is strongly affected by the small number of events.

In order to evaluate the effects of subsequent ionization steps on the final ion momentum, coincidence detection of ion and emitted electrons has been used, see e.g.~\cite{Ulrich2003,Pfeiffer2011}. However, efficient detection of more than two emitted electrons per laser shot is very challenging. Therefore, we propose and utilize an alternative scheme based on the idea that the final ion momentum is the convolution of the momentum distributions of all ejected electrons. For double ionization, e.g., the first ionization step from Ne$^{1+}$ to Ne$^{2+}$ would be described by a distribution $F_{\rm single,1}$ and the second ionization step from Ne$^{2+}$ to Ne$^{3+}$ by  $F_{\rm single,2}$, each parameterized by the four quantities found in eq.~(\ref{eq:gaus}). The resulting distribution for the sequential double ionization $F_{\rm double}$ will be the convolution $F_{\rm double}=F_{\rm single,1} \ast F_{\rm single,2}$.

Accordingly, a function of the form of $F_{\rm double}$ is fitted to the measured Ne$^{3+}$ momentum distributions. This procedure yields the set of parameters defined in equation~(\ref{eq:gaus}) for each individual ionization step. The success of the approach is demonstrated in Fig.~\ref{fig:Deconvolution} for the example of double ionization of Ne$^+$ with $\epsilon=$0.74 as shown in Fig.~\ref{fig:dis_neon}. The procedure was also tested with momentum distributions generated by Monte-Carlo simulation. We found that the fit results for $p_{r0}^{1\rightarrow 2}$ and $p_{r0}^{2\rightarrow 3}$ deviate less than 1\% from the simulation values in the case of double ionization. This increases for more ionization steps to not larger than 7\% in the case of quadruple ionization, see Fig.~3b).

Thus, the deconvolution procedure yields precise $p_{r0}$ values for every ionization step~(see table \ref{table_element}). We label the individual ionization steps from charge state $n$ to charge state $n+1$ with tuples $nk$, where $k$ is the charge state in which the ion will eventually end up. For example, '25' corresponds to the distribution for ionization of Ne$^{2+}$ to Ne$^{3+}$ for ions that will subsequently further ionized to Ne$^{5+}$.

\begin{figure}[tbh]
\includegraphics[width=1\columnwidth]{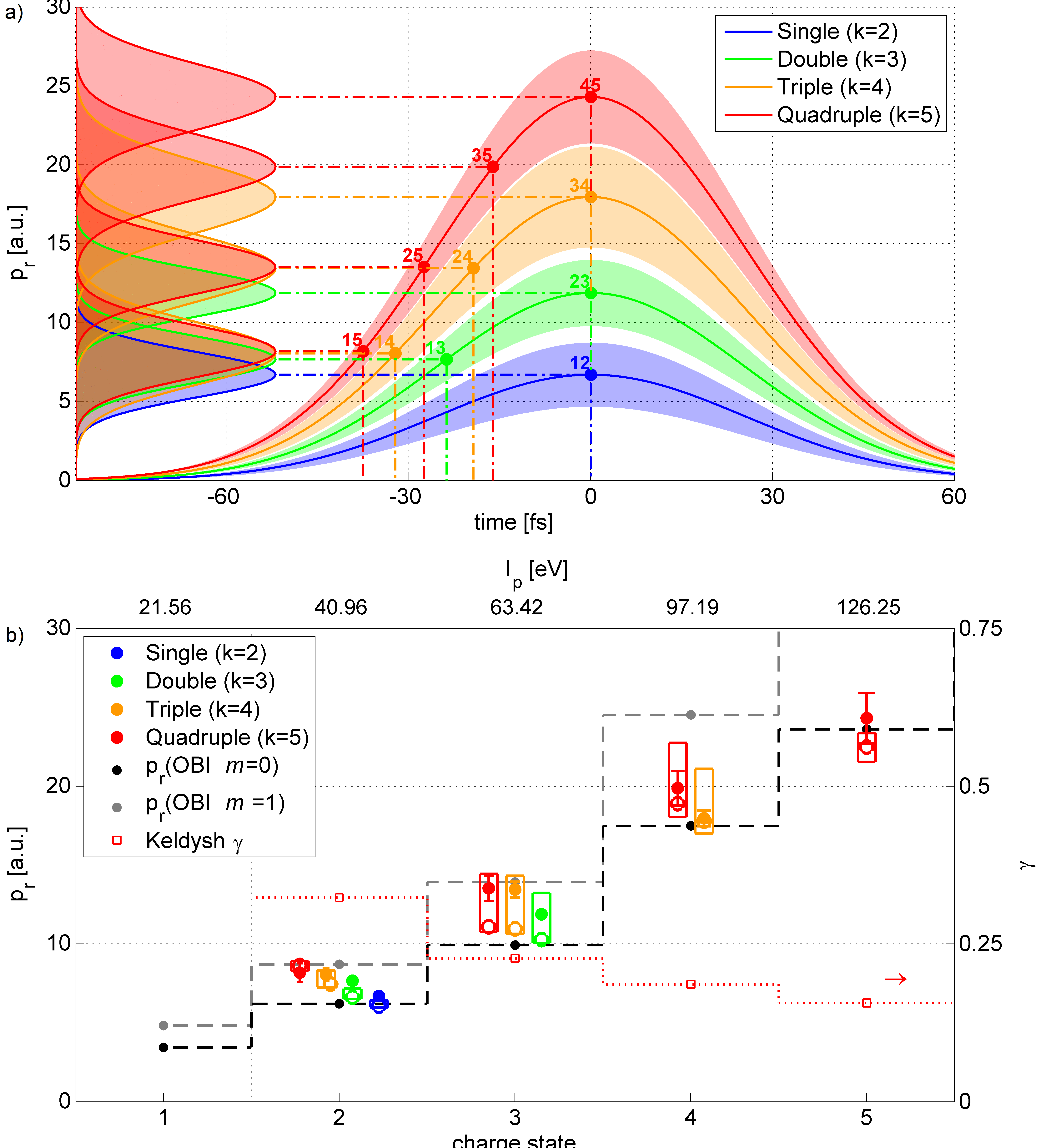}
\caption{(color online). a) The radial electron momentum $p_{r0}$ along with its FWHM $\Delta p_{r0}$ are depicted in vertical direction. The time of ionization can be obtained by projection on the envelope of the vector potential in the respective focal volume. b) $p_{r0}$ as function of the charge state $n$ of the respective individual ionization event for the different final charge states $k$. The solid circles show measured values, where the color refers to the charge state $k$ in which the ion eventually ends up. For visual convenience, the data points are displaced in horizontal direction. Open circles and rectangles are the results of ADK calculations, where focal volume averaging has been taken into account (peak intensity in the simulation is $1.1\cdot10^{17}\,\Wcm$). The open circles show the result of the simulation, where an equal distribution among the available m quantum numbers has been assumed. The rectangles enclose the range given by the two extremal distributions of the $m$ quantum number. The black and the grey data points refer to the radial momenta following from the formula  for the OBI for $m$=0 and $m$=1. The red open squares (right scale) represent the trend of Keldysh parameter.}
\label{fig:timing}
\end{figure}

For the intensities available in the experiment, Ne$^+$ can be ionized up to Ne$^{5+}$. Accordingly, for Ne$^+$ the ionization processes '12', '13', '14', and '15' exist, each with a value for $p_{r0}$ (as well as the other three parameters). For the further discussion, it is important to realize that for each but the last $(n=k-1)$ ionization step, $p_{r0}$ corresponds to the respective saturation intensity due to the highly nonlinear dependence of ionization on intensity. This can be confirmed experimentally by studying the intensity dependence of $p_{r0}$: when the saturation intensity is reached, $p_{r0}$ will stay constant. Process '12', e.g., will take place only in the outer regions of the laser focus, while '15' is only possible in its center. 

\begin{table}[tbh]
\begin{center}
\begin{tabular}{ccc||c|c}
$\epsilon$  & Transition & Final & $p_{r0}$ [a.u.]  & $I_{pr0}$ [W$/$cm$^2$] \\ \hline 
0.74 & 1$\rightarrow$2 & 2 &   6.703$\pm$0.05 &   (5.1$\pm$0.07)$\cdot$10$^{15}$\\ 
0.74 & 1$\rightarrow$2 & 3 &   7.67$\pm$0.15  &   (6.7$\pm$0.3)$\cdot$10$^{15}$ \\
0.74 & 1$\rightarrow$2 & 4 &   8.05$\pm$0.4   &   (7.4$\pm$0.7)$\cdot$10$^{15}$ \\
0.74 & 1$\rightarrow$2 & 5 &   8.18$\pm$0.6   &   (7.6$\pm$0.9)$\cdot$10$^{15}$ \\
0.74 & 2$\rightarrow$3 & 3 &   11.88$\pm$0.15  &  (1.61$\pm$0.04)$\cdot$10$^{16}$ \\
0.74 & 2$\rightarrow$3 & 4 &   13.45$\pm$0.5  &   (2.06$\pm$0.15)$\cdot$10$^{16}$ \\
0.74 & 2$\rightarrow$3 & 5 &   13.53$\pm$0.8  &   (2.09$\pm$0.24)$\cdot$10$^{16}$ \\
0.74 & 3$\rightarrow$4 & 4 &   17.96$\pm$0.5  &   (3.68$\pm$0.20)$\cdot$10$^{16}$ \\
0.74 & 3$\rightarrow$4 & 5 &   19.87$\pm$1.1  &   (4.50$\pm$0.48)$\cdot$10$^{16}$ \\
0.74 & 4$\rightarrow$5 & 5 &   24.3$\pm$1.6   &   (6.73$\pm$0.85)$\cdot$10$^{16}$ \\ \hline

\end{tabular}
\end{center}
\caption{Fit parameters and the associated intensities from the results of the deconvolution method presented in Fig.~\ref{fig:Deconvolution}. The ellipticity was fixed to the experimentally measured value.}
\label{table_element}
\end{table} 

For a well-characterized laser pulse $p_{r0}(nk)$ can be used to determine the space of time, $t_0\pm\Delta t_0$, during which the respective ionization process took place. This is depicted in Fig.~\ref{fig:timing}a) for the pulse duration of 36\,fs used in the experiment. In Fig.~\ref{fig:timing}b) the various values for $p_{r0}(nk)$ are plotted as a function of the respective charge state $n$ (solid circles).

Fig.~\ref{fig:timing} shows that the same ionization step $n \rightarrow n+1$ results in a slightly different momentum $p_{r0}$ depending on the final charge state $k$. We observe the consistent trend that events with equal $n$ occur at the higher intensities the larger $k$. This can be explained by inspection of Fig.~\ref{fig:timing}a). For higher peak intensities -- which are reached closer to the focus -- ionization of low charge states shift towards the beginning of the laser pulse. For the present conditions, this implies ionization on a steeper pulse slope and thus a higher saturation intensity. The fact that the deconvolution procedure is accurate enough to detect the dependence of the saturation intensity on the pulse envelope is remarkable. 

Next, we compare our measured results with predictions of the frequently used model that ionization saturates when the wavepacket can pass the Coulomb potential over the barrier (OBI), i.e. when the peak intensity is greater than $I_{\rm OBI}$. In a one-dimensional model, this intensity is given by $I_{\rm OBI}^{\mathrm{1D}} =  \Eip ^{4}/(16Z^2)$ ~\cite{Augst1989}. From Fig.~\ref{fig:timing}b) it is obvious that the OBI formula calculated in 1D predicts considerably too low ionization intensities (see black curve in \ref{fig:timing}b)).

Better quantitative agreement can be obtained with classical Monte Carlo simulations when the ionization rate is given by the ADK formula. Averaging over the 3D focal intensity distribution and all possible carrier envelope phases is taken into account, while the influence of the Coulomb potential is neglected. The ionization rate and therefore the result for the predicted momentum depends on the absolute value of the $m$ quantum number. Corresponding results for $p_{r0}$ are plotted in Fig.~\ref{fig:timing}b) (open circles) assuming an equal distribution of the possible $m$-states~\cite{Pfeiffer2011a}. In addition the limiting cases for the $m$ number distribution are marked by the rectangles. The upper limit of the momentum corresponds to the minimum number of electrons with $|m|=0$ and the lower limit to the case of maximum number in $|m|=0$.

The ADK based simulation with averaged $m$ quantum number underestimates the final radial momentum systematically, which may be explained by the well-known fact that the ADK rate is too high for the conditions of this experiment~\cite{Scrinzi1999}. Nevertheless the limit of the lowest ionization rate corresponding to the case with minimum number of electrons with $|m|=0$ results in a broad possible momentum range, which includes the measured momenta. For comparison in Fig.\ref{fig:timing}b) (grey line) the modification of the OBI for $|m|$=1 is shown~\cite{Shakeshaft1990}. For a correct modeling of the ionization steps the production of coherent electron wavepackets inside the remaining ion has to be considered in each ionization process \cite{Rohringer2009,Worner2011,Pfeiffer2013}, as the time-dependent orientation of the electron holes in the the valence shell will influence the ionization rate of subsequent ionization events in a complex way. Therefore, for modeling plasma dynamics, no simple formulae are reliable, which suggests that measured data should be used for accurate calculations.

In conclusion, multiple ionization up to quadruple ionization of Ne$^{1+}$-ions by elliptically polarized laser pulses was investigated. We introduce a method to deconvolve the measured momentum distribution of multiply ionized ions and extract the saturation intensities from the ion momenta for each single ionization step in the multiple sequential ionization process. Thereby, removing the typically large experimental uncertainties in the intensity determination, this enables us to track the averaged ionization times of all charge states created in the laser focus. The method is sensitive enough to detect the dependence of the saturation intensity on the slope of the pulse. The theoretical model that resulted in the best agreement with the experimental data is based on the ADK rate. The method can also be applied to other wavelengths as long as $\gamma$ remains small. Longer wavelength, in particular, will increase accuracy. For few-cycle pulses, the influence of the carrier-envelope phase needs to be taken into account. Another option is the extension to higher intensities by using ion beams with higher initial charge state. Thus, the method can be used to verify the modeling of the ionization dynamics, which is the basis of all strong-field laser matter interaction, for example in plasma physics.

This work was supported by grant PA730/4 of the German Research Foundation (DFG).

\bibliographystyle{apsrev4-1}
\bibliography{library2}


\end{document}